| | |
|---|---|
| Title | **Integrated micro-plasmas in silicon operating in helium** |
| Authors | R. Dussart[1], L.J. Overzet[2], P. Lefaucheux[1], T. Dufour[1], M. Kulsreshath[1], M.A. Mandra[2], T. Tillocher[1], O. Aubry[1], S. Dozias[1], P. Ranson[1], J.B. Lee[2], and M. Goeckner[2] |
| Affiliations | [1] GREMI, CNRS/University of Orleans, 14 rue d'Issoudun, BP 6744, 45067 Orléans Cedex 2, France<br>[2] Plasma Science and Applications Laboratory, University of Texas at Dallas, 800 W. Campbell Road, RL10, Richardson TX 75080-3021, USA |
| Ref. | The European Physical Journal D, 2010, Vol. 60, Issue 3, 601-608 |
| DOI | http://dx.doi.org/10.1140/epjd/e2010-00272-7 |
| Abstract | Microplasma arrays operating in helium in a DC regime have been produced in silicon microreactors. Cathode boundary layer (CBL) type microdevices were elaborated using clean room facilities and semiconductor processing techniques. Ignition of the micro-discharge arrays having either 50 or 100 μm diameter cavities was studied. Two different structures (isotropically etched or anisotropically etched cavity) and various conditions (the two different voltage polarities, pressures etc.) were investigated. 100 microdischarges of 50 μm diameter could be ignited in parallel at 1000 torr. At high current, some parasitic and transient sparks appeared at the edge of the sample. When the polarization was reversed (cathode side corresponding the opened electrode), more current was needed to light all the microdischarges. A thermally affected zone around the hole on the anode side was obtained after operation. |

# 1. Introduction

Micro hollow cathode discharges (MHCDs) were first introduced by Schoenbach et al. in 1996 [1]. The structure of MHCD reactors tends to be quite simple: they generally consist of two electrodes separated by an insulator, much like a capacitor. The difference is that this structure has at least one and sometimes many holes drilled through it. By applying a high voltage (typically several hundreds of volts) between the two electrodes, one can initiate a stable glow discharge at atmospheric pressures localized inside the hole(s). The term MHCD for this type of discharge was adopted by several groups [2–7] before Boeuf et al. [8] showed that the structure was not due to the classical hollow cathode effect. Nevertheless, the term MHCD was maintained in the community. The typical V-I curve of a MHCD can be divided into 3 conceptual parts: (1) at small current levels, the voltage increases with current and the discharge remains inside the hole. (2) At large current levels, the voltage remains more or less constant as the current increases and the plasma extent over the cathode surface increases with current. This is the so-called "normal glow" regime. In the typical normal glow regime, several watts can be injected into each MHCD corresponding to several hundreds of kW cm$^{-3}$. The electron density can reach $10^{15}$ cm$^{-3}$ [9]. (3) At intermediate currents, a self-pulsing regime can be obtained where there are large current fluctuations and the plasma oscillates between remaining inside the hole and extending over the cathode surface [10,11].

MHCD reactors are often assembled mechanically and drilled either with a laser beam or mechanically. In these cases the diameter of the MHCD reactor is large, typically greater than 100 μm. Moreover, the thickness of the metal electrodes is typically similar to the thickness of the dielectric. This can limit the production and functionality of the devices. Eden's team was the first to use clean room techniques to fabricate arrays of MHCD reactors on a silicon substrate [12]. The typical dimensions of those devices were 50×50 μm². He used silicon nitride layers and polyimide as the dielectrics. The cathode was made by wet etching the silicon substrate to form inverted pyramids. The anode was made by depositing nickel. As a result, they were able to ignite up to 250 000 micro-discharges all together using AC excitation [13].

Our own research into MHCDs is a collaboration between colleagues at GREMI (Orleans, France) and the University of Texas at Dallas (UTD, Texas USA) first started in 2006. The first samples were partially prepared in the clean room of the engineering school at UTD. Relatively thick alumina substrates ($Al_2O_3$, ~500 µm) were electroplated with a thin layer of nickel (~7 to 8 µm) and laser drilled to form either one or multiple MHCD reactors in parallel in various hole arrangements. They were tested at both labs. At GREMI, tests were made in He at pressures between 20 and 1000 torr using a single ballast resistor for the entire array. It was found that although a single hole ignited readily, lighting multiple holes simultaneously was much more difficult. In fact, the first hole to ignite would self-destruct from being over-powered before a second would ignite. This is unfortunate since arrays of MHCD reactors in parallel can enable many more applications than isolated MHCD reactors. It allows one to extend the micrometric scale of one MHCD to the macroscopic scale of an array for plasma processing applications. Other research groups were able to ignite multiple microdischarges mounted in parallel. Of those reports, some had to use an individual ballast resistor for each MHCD [14,15] while others did not [16,17]. The reason for this was not clear until recently when we discovered why individual ballasts are required in some configurations and not necessary in some others [18]. We showed that limiting the cathode surface area allows one to obtain an abnormal glow regime of the MHCD and ignite all of the microplasmas in an array.

This preliminary work proved useful for designing our newer integrated devices in silicon. At atmospheric pressure, more radicals can be formed for use in materials or gas processing. The goal is to well control the operation of each micro-discharge in the array so that processing at the macroscopic scale can be optimized too. It turns out that silicon is a rather good material for such reactors because it has good mechanical and thermal properties (fusion temperature of 1414°C, good thermal conductivity of 149 $Wm^{-1}.K^{-1}$, linear thermal expansion coefficient of $2.6 \times 10^{-6}$ $K^{-1}$ ...). Some effects due to its being a semiconductor can also be expected as was shown by Ostrom and Eden [19].

Our objective is to fabricate and study integrated micro-plasmas in silicon to investigate their limits and characteristics for their future implementation in systems: lab on chip, micro electro mechanical systems (MEMS), micro-sensors, etc. Whereas micro-fluidic devices are already used extensively in MEMS devices, micro-plasmas are not yet typically found. Before their implementation, micro-plasmas should be studied to find their limits in terms of dimensions, lifetime, and physical characteristics (density, temperature etc.). Such studies can be begun by electrical and optical characterization. Here, we re- report some first experiments using silicon micro-reactors, which were fabricated as a collaboration between UTD and GREMI using semiconductor processing techniques. MHCD arrays consisting of 100 holes having either 50 or 100 µm diameters were made for these initial tests although much smaller devices are possible for future generations. The micro-fabrication sequence will be described in Section 2. Ignition of the micro-discharge arrays operating in helium will be discussed in Section 3 for two different structures (isotropically etched or anisotropically etched cavity) and various conditions (the two different voltage polarities, pressures etc.).

## 2. Experimental setup & micro-reactor fabrication

As noted, MHCD reactors were fabricated on silicon wafers using standard clean room processes. We used 3" n-type silicon wafers for these first tests and designed so called "cathode boundary layer" (CBL) discharges [20]. A CBL discharge is simply a high-pressure glow between a planar cathode and ring-shaped anode that are separated by a dielectric. Typically the dielectric thickness has been of the order of 100 µm [20]; however, we chose to use a $SiO_2$ dielectric thickness of 5 µm since we used a low pressure chemical vapor deposition (LPCVD) technique to deposit it. Growing a much thicker oxide is time and cost prohibitive. This thickness of oxide is easily sufficient to withstand the electric fields required for discharge breakdown. Different geometries were designed on the wafer to investigate micro-discharges having different dimensions. Both single hole and 10×10 arrays having either 50, 100, 150 or 200 μm diameters were fabricated.

The process flow is illustrated in Figure 1 with Figure 1.4c showing the top view and Figures 1.1–1.4b showing the A-A cross sections. The 5 μm thick $SiO_2$ layer, represented in white in Figure 1, was grown using an LPCVD process (Fig. 1.1) at UTD. The oxide on the backside of the wafer was stripped in buffered oxide etch solution while the front side was protected using photoresist. On top of this front-side oxide, a thin adhesion and nucleation layer of Cr/Cu was evaporated (not represented in Fig. 1). Subsequently, the first lithography step was used to pattern the MHCD holes and define the overall anode area (resist is represented in orange in Figs. 1.2 and 1.3). An electro-deposition process on the unprotected areas of the Cr/Cu film was used to grow the nickel anode layer. This Ni film was made to be a few microns thick and is designated by the green film in Figures 1.2–1.4. The nickel layer was left thinner than what is typically used for MHCD reactors [20] because of a limitation set by the photoresist thickness. A second lithography step was carried out to protect the $SiO_2$ layer surrounding the anode during etching of the oxide layer in the hole (not represented in Fig. 1). Maintaining the integrity of the surrounding insulator is thought to help avoid arcing at the wafer edges (between nickel anode and silicon cathode in Fig. 1.3). As we will note later, this oxide layer is not sufficient to prevent arcing at the edges. The $SiO_2$ in the holes was then etched using a reactive ion etch (RIE) process at GREMI. The Ni anode was used as a hard mask for that etch step. The penultimate step was plasma etch of the silicon substrate itself to define the cathode area and shape. Two different processes were utilized to either form an anisotropic etch profile (Fig. 1.4a) or a more isotropic etch profile of the silicon (Fig. 1.4b). The final step was to deposit aluminum on the back side of the wafer to provide an ohmic contact to the cathode (Si). This is the green film in Figures 1.4a and 1b.

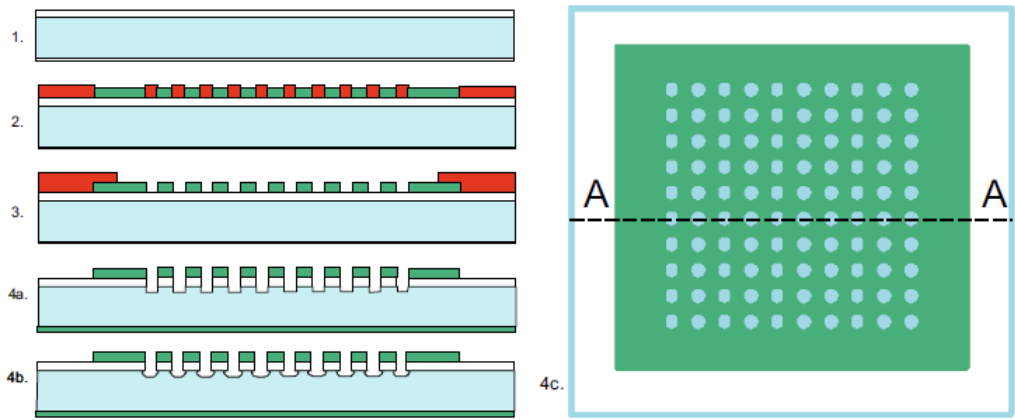

*Fig. 1. Schematics illustrations of the wafer cross section after the various process steps used to form a 10 × 10 array of MHCDs on a silicon substrate. (1) After forming the $SiO_2$ dielectric. (2) After electro-plating the Ni anode. (3) After developing the photoresist for $SiO_2$ protection. (4a, 4b) Final state after etching the Si and depositing the backside electrical contact. (4c) Top view.*

Figure 2 is a scanning electron microscope (SEM) image showing 150 μm diameter micro-reactors after operation in helium. The holes were etched isotropically as drawn in Figure 1.4b. The nickel and $SiO_2$ layers are clearly visible at the top of the structure (the oxide protrudes past the silicon sidewall) and the silicon cavity is ~60 μm in depth. The SEM image in Figure 3 shows 100 μm diameter MHCD reactors that were etched anisotropically to form 10 μm deep holes. The two cases are quite different: the distance between the two electrodes is much greater for the isotropically etched MHCD reactors. Moreover, the isotropic profiles of the 150 μm diameters cavities provide a much greater area than the anisotropically etched 100 µm diameter cavities. As mentioned in the introduction, the total area of the cavity is relevant if it is used as a cathode: it will define the maximum current which can be driven in each cavity.

In order to make the first tests, we placed the chip into a vacuum system which could be filled with helium at a chosen pressure. A very slow triangle wave voltage signal (0.025 Hz) was applied using a 40 kΩ ballast resistance and the discharge voltage and current were recorded. Both discharge ignition and the appearance of the discharge(s) were observed for 1 and 100 hole MHCD at varying values of the current.

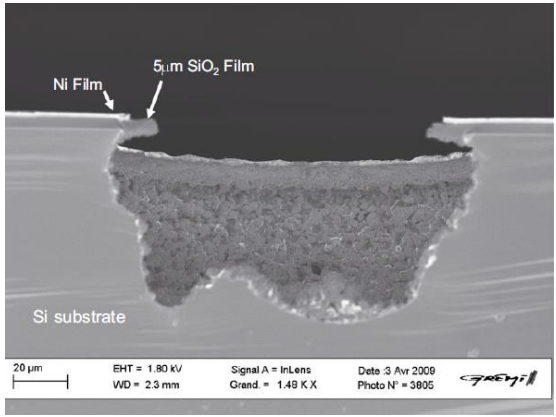 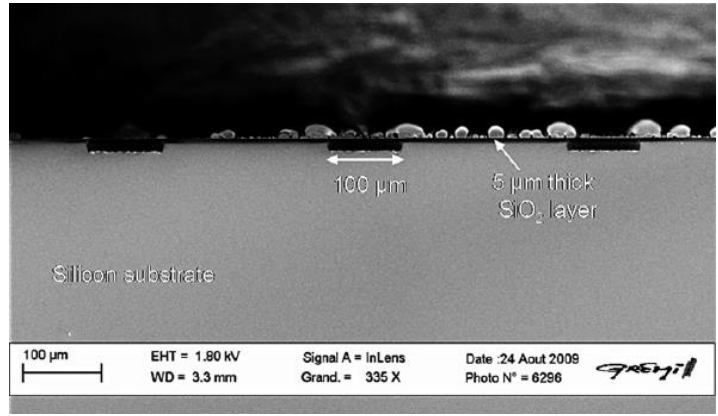

*Fig. 2. MHCD hole profile obtained after an isotropic plasma etching process showing the nickel/SiO$_2$/silicon structure.*

*Fig. 3. MHCD hole profiles obtained after an anisotropic plasma etching process showing the nickel/SiO2/silicon structure. Note that the nickel layer does not appear on this SEM picture.*

## 3. Results & Discussion

We first show the results obtained for a single hole of 150 µm diameter. Figure 4 shows a typical V-I curve for an anisotropically etched MHCD sample operating in helium at 350 torr. The distance between the cathode and anode was ~5 µm (the dielectric thickness). The cathode area (the Si surface area) was also smaller than that of samples etched isotropically. It was estimated to be ~0.018 mm$^2$. The current was found to be of the order of 100 µA as shown in the V-I characteristic of Figure 4. The left inset is a picture of the micro-discharge for a current below 100 µA. Beyond 250 µA, the micro-discharge entered an abnormal glow regime. As observed on the V-I characteristic and on the right inset, several parasitic discharges ignited at this higher current level. The exposure time for each picture was 0.1 s, so that each picture also integrated transient (pulsed) discharges, which are clearly evidenced in the V-I characteristic as the sudden transitions to large currents and subsequent return to the main V-I characteristic. At small currents, 2 parasitic microdischarges are in the proximity of the bright main discharge. These are attributed to nickel voids formed during the electro-plating step. Such defects can allow an undesired SiO2 etch and the formation of unwanted cavities in these locations. They can be removed by depositing some epoxy in their locations as shown next.

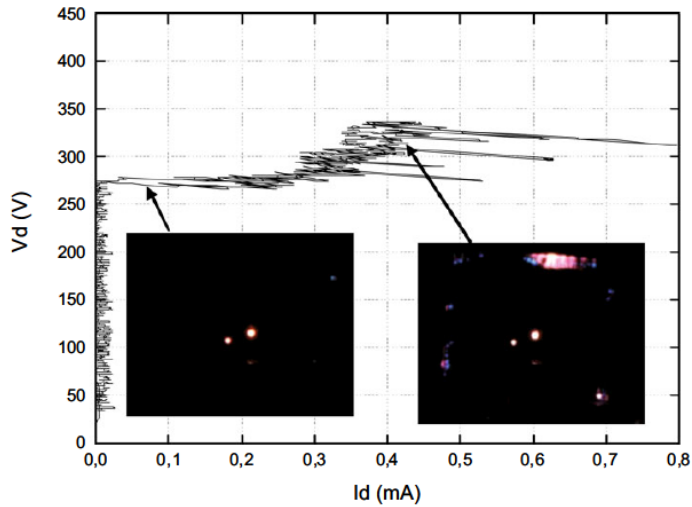

Fig. 4. V -I curve obtained for a single hole micro-discharge working in helium at 350 torr. Insets: pictures of the micro-discharges for a current discharge of 100 µA (left) and for 400 µA (right).

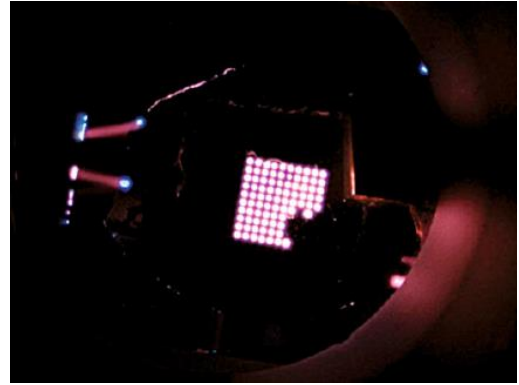

Fig. 5. Picture of a 10×10 micro-discharge array (100 µm diameter) operating in helium at 500 torr for a total current of around 20 mA showing the main micro-discharges and some transient sparks which appear at the edges of the sample.

Figure 5 is a picture (200 ms exposure) of a 10 × 10 array of MHCDs having diameters of 100 µm and operating in helium at 500 torr. The silicon surface was biased to be the cathode of the micro-discharge. Only 83 of the 100 micro-reactors, could ignite because the 17 holes at the bottom right were accidently covered by epoxy during sample preparation. We note that these MHCD samples were etched isotropically, forming cavities like the one shown in Figure 2. As a consequence, the current flowing through each micro-discharge was somewhat larger than that found in the case of the single micro-discharge presented in the previous paragraph (Fig. 4). In fact, currents as large as 26 mA were obtained for this sample corresponding to an average of ~300 µA per MHCD. We estimate that the cathode area for each micro-cavity is at least equal to 0.065 mm$^2$, which is ~3.5 times more than the cathode area of the single cavity sample (ignoring surface area enhancement due to the significant surface roughness seen in Fig. 2). Although the available cathode area of individual micro-discharges is larger, some parasitic current flow is observed going to the edge of the sample in Figure 5. In particular, two plasma filaments appear on the left side of the picture between silicon (far left edge) and nickel anode. Some other sparks can also be observed in the picture. Again, those parasitic discharges are not stable, but rather transient sparks, which are characterized by current and voltage spikes on the V -I characteristic.

Such spikes can be observed in a time-dependent plot of the current shown in (Fig. 6a) recorded by our oscilloscope. Short spikes in both the current and voltage were found to be numerous especially when the average current exceeded 10 mA. The V -I curve can be plotted by removing these spikes to enhance its readability (Fig. 6b). One can see that the discharge voltage remains ~200 V at these experimental conditions. While the discharge voltage rises a small amount with the current above 15 mA, no clear indication of entering the abnormal glow regime was observed for this sample from 2 to 27 mA.

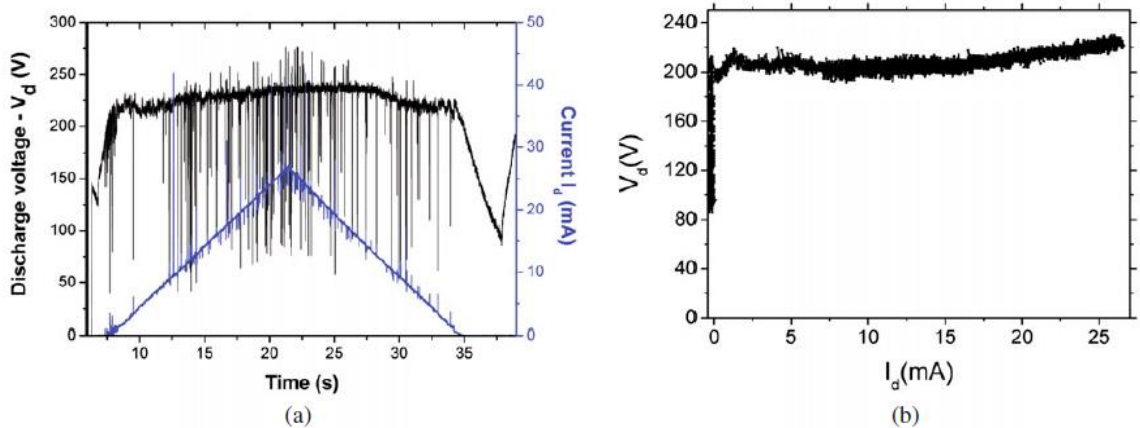

*Fig. 6. (a) Discharge voltage and current versus time for a 10 × 10 micro-discharge array (100 μm diameter) operating in helium at 500 torr. The current follows the triangle wave, while the discharge voltage saturates. (b) Resulting V -I characteristic when current and voltage spikes due to parasitic discharges were removed.*

To better understand the ignition sequencing of the micro-discharges, a picture of the sample was taken every 3 s during the increase and decrease of the current (as shown in Fig. 6). The results are presented in Figure 7, which is a sequence of 7 pictures in time order starting at small currents (Fig. 7a) rising to the maximum current (Fig. 7d) and then falling back toward extinction (Figs. 7e–7g). Figures 7a and 7b show the device at currents between 3 and 8 mA. The cavities in the upper right and lower left portions of the sample ignite first and as the current rises, more micro-discharges ignite. By 10 mA (Fig. 7c), all 83 micro-discharges were ignited. At currents larger than 10 mA, sparks appeared at the edge of the array (Figs. 7d and 7e). As the current decreased (Figs. 7f and 7g), the extinction sequence followed the same trend in reverse. This experiment shows first that the microdischarges of the array do not all ignite simultaneously at a single breakdown voltage. The current limitation set by the external ballast resistor limits the area of cathode which can be used in forming the glow and thus acts to limit the number of holes which can be ignited at any given moment. A related phenomenon was reported by Waskoenig et al. for a similar type of micro-discharge, but using an AC power supply at high frequency [21]. This results in an ignition sequence for holes in the array which appears to occur in reverse during the current decrease. Finally, transient sparks appear at the edge of the sample at high current levels once all the MHCD holes are ignited. It could be that the DC plasma finds additional cathode area in this fashion.

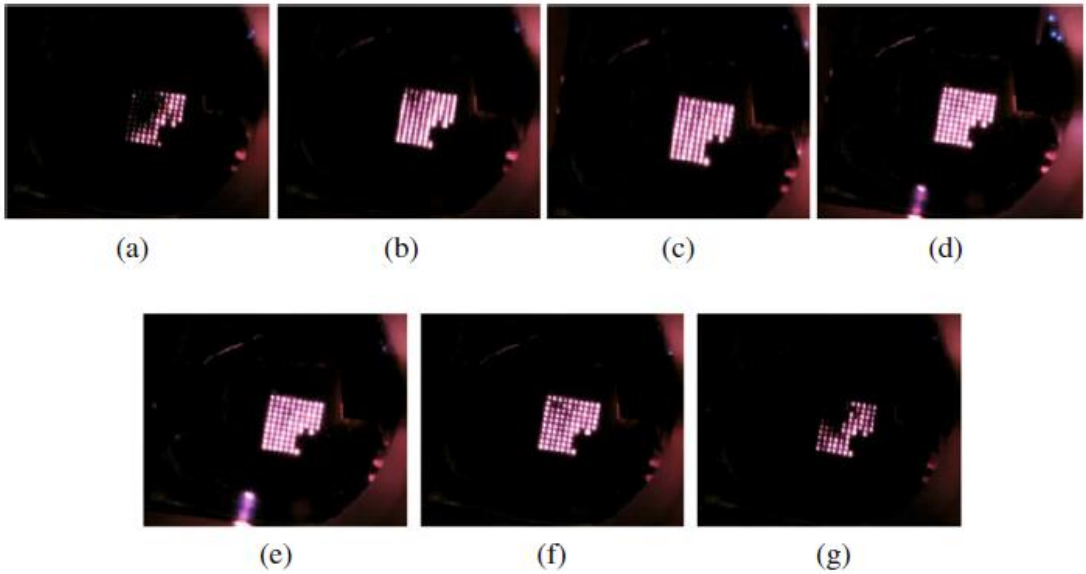

*Fig. 7. Temporal evolution of the micro-discharge array during current increase (a–d) and decrease (e–g).*

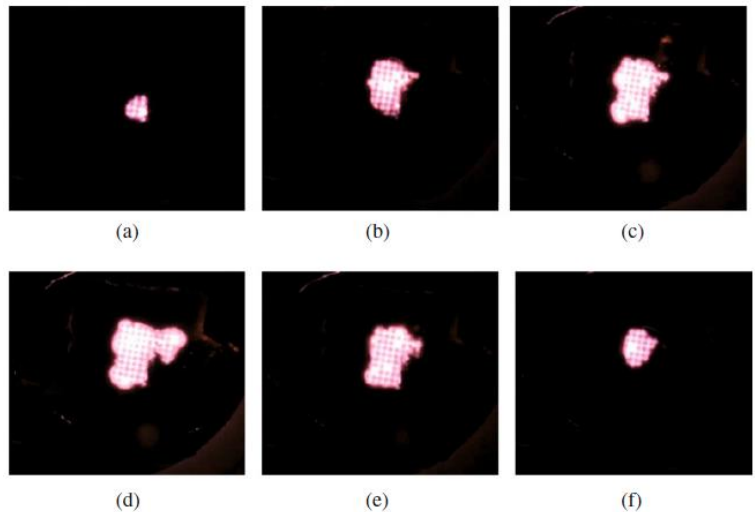

*Fig. 8. Temporal evolution of the micro-discharge array during current increase (a–d) and decrease (e–f) when the nickel is biased to be the cathode.*

If we invert the polarity of the voltage applied to the sample, then the nickel film becomes the cathode, and the silicon becomes the anode for the micro-discharges. In this case, the MHCDs can still ignite, but take on a very different appearance and characteristic (see Fig. 8. where a sequence of pictures analogous to those in Fig. 7 is shown). Note that the plasma did not appear connected to the cavities or to significantly enter the cavities, but simply expanded over the cathode area instead. As the current increased, the plasma extended over a larger surface area but never covered the full array. Indeed, when the nickel was the cathode, the cathode area was not as effectively limited and as a consequence the current could increase without reaching an abnormal glow regime. The result is that larger currents were required to form a plasma having a similar spatial extent as that for the array with the Si as the cathode.

We also ignited micro-discharges at higher pressure (1000 torr) using the same sample. A picture of the microdischarge array is shown in Figure 9. Even at larger current levels, all the micro-discharge cavities did not ignite. Moreover, among the ignited micro-plasmas, emission intensity was not uniform: some were very bright while others were weak. As seen on the picture, we did not observe any spark at the edge of the sample in these conditions even though the current level was well above that producing sparks at lower pressures.

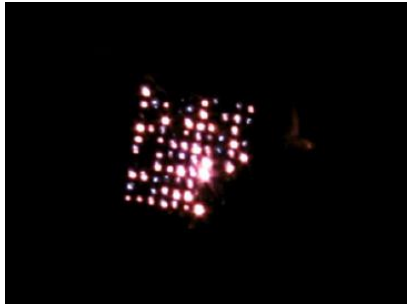

*Fig. 9. Micro-discharge array at 1000 torr in helium (100 μm diameter cavities) at 20 mA.*

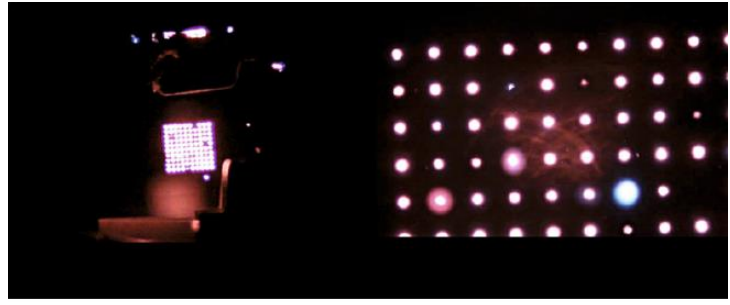

*Fig. 10. (a) Picture of a 10 × 10 micro-discharge array (50 μm diameter) operating in helium at 1000 torr for a total current of ~15 mA showing the main micro-discharges and some transient sparks which appear at the edges of the sample. (b) Picture taken in the same experimental conditions, but with a higher optical magnification.*

Experiments were also carried out on smaller diameter cavities, having a shape like the one of Figure 3. We expect that the inter-electrode distance was reduced compared to that in isotropically etched cavities. Despite this, it was possible to ignite MHCDs as shown in Figure 10a for 50 μm diameter cavities. Figure 10b is another picture of the micro-discharges made in the same experimental conditions, but at a higher magnification. Interestingly, higher pressures were required to ignite the discharge. Figure 10 shows the results at 1000 torr, the pressure used in Figure 9 which resulted in incomplete ignition of the array. For the anisotropically etched array, microplasmas did not ignite at lower pressures, even at 700 torr; but nearly all the micro-discharges ignited at 1000 torr. At the larger current levels for this array some parasitic sparks were observed at the edge of the sample (Fig. 11a). Among the ignited micro-discharges, one can notice that the emission intensity appears inhomogeneous, which indicates that the current is not equally distributed between the different micro-discharges. Although we do not have yet a definitive explanation for it, different interpretations can be given: first, holes are not exactly the same in terms of electrode roughness and some of them could be more efficient to drive current; second, the exposure time of the camera being quite long (100 ms), discharges might not be as stable as they seem to be and some of them could appear brighter than others.

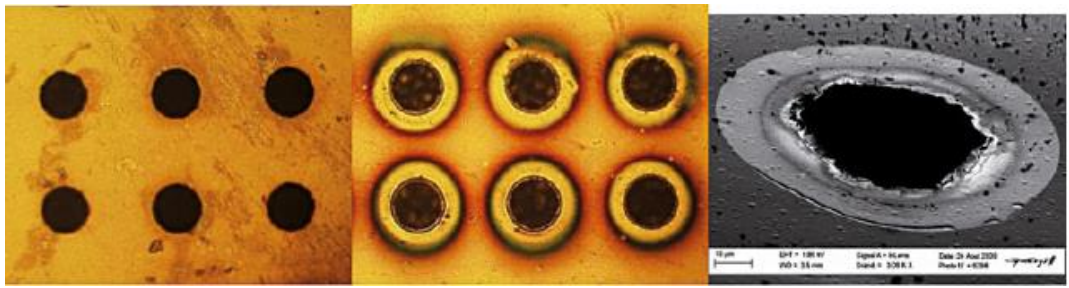

*Fig. 11. (a) Optical microscope picture of the Ni film before the ignition of a micro-discharge. (b) Optical microscope picture of the Ni film on the same sample after the ignition of micro-discharges. (c) SEM image of the Ni film showing erosion after operation.*

In Figure 11, we show some optical and SEM images of the Ni film before (Fig. 11a) and after (Figs. 11b, 11c) operation. A 20 micron wide ring appears around the cavity after operation even though the Ni film is biased as the anode. The edge of the film also becomes rough at the microscopic scale (Fig. 11c). If we assume that all the microdischarges in the array ignited and that the total current lay between 1 and 10 mA, we can make a rough estimation of the current density at the anode. For a 50 µm diameter cavity, if we were to assume that all of the electrons entered the Ni (anode) film solely through the 5 micron thickness, then the current density would have to be quite large. It would lie between 1.3 and 13 $Acm^{-2}$. Therefore, it is probable that a fraction of the electron current to the anode is injected on the top surface and should be the origin of the observed 20 µm wide ring. Given this, the required current density is reduced to a value between 0.2 and 2 $A.cm^{-2}$, a more reasonable range. As a consequence, in our devices with a 5 µm thin metal layer for the anode, we can think that a part of the current is driven through a larger ring around the cavity opening. This ring could become heated and as a result, eroded and roughened. It may also be that the metal evaporates at this location. Optical emission spectroscopy should confirm this and we plan to carry out the experiment in the near future.

## 4. Conclusions

In this paper, we have presented some first results obtained on silicon CBL type micro-reactors having two different types of geometry: one formed an isotropic cavity and the other one had a smaller area and cylinder cavity. A significant part of this work was dedicated to the microfabrication of the devices. We have shown that microdischarge arrays operating in helium could be obtained at different pressures. At high current, some parasitic and transient sparks were obtained. When the polarization was reversed, we could also ignite micro-discharges, but more current was needed to light them all. At high pressure, the intensity of the micro-discharges was not homogeneous on the array. Finally, by analyzing our samples by SEM and optical microscopy after operation, we could observe a thermally affected zone around the hole on the anode side. This ring shape region seems to correspond to the area from which the current is driven through around the cavity opening. The samples we used were quite fragile and had a limited life time (about 10 min for an injected current between 1 and 20 mA). After several tests, a short circuit appeared, so that no micro-plasma could be obtained anymore. New samples are being fabricated, taking into account these first results.

This work is financially supported by the French "Agence Nationale de la Recherche" through the contract No. ANR-09- JCJC-0007-01 under the name SIMPAS project.